\documentclass{iau}
\usepackage{graphicx}

\title[DESK Survey Results] 
{First Results from the Disk Eclipse Search with KELT (DESK) Survey}

\author[Joseph E. Rodriguez$^{1}$, Joshua Pepper$^{2,1}$, Keivan G. Stassun$^{1,3}$]   
{Joseph E. Rodriguez$^{1}$,
 Joshua Pepper$^{2,1}$ 
 and  Keivan G. Stassun$^{1,3}$}

\affiliation{$^1$Department of Physics and Astronomy, Vanderbilt University, 6301 Stevenson Center, Nashville, TN 37235, USA\\email: {\tt rodriguez.jr.joey@gmail.com}\\
$^2$Department of Physics, Lehigh University, 16 Memorial Drive East, Bethlehem, PA 18015, USA\\email: {\tt joshua.pepper@Lehigh.EDU}\\
$^3$Department of Physics, Fisk University, 1000 17th Avenue North, Nashville, TN 37208, USA\\email: {\tt keivan.stassun@vanderbilt.edu}}

\pubyear{2015}
\volume{314}  
\pagerange{119--126}
\jname{Young Stars \& Planets Near the Sun}
\editors{J. H. Kastner, B. Stelzer, \& S. A. Metchev, eds.}
\begin{document}

\maketitle

\begin{abstract}
Using time-series photometry from the Kilodegree Extremely Little Telescope (KELT) exoplanet survey, we are looking for eclipses of stars by their protoplanetary disks, specifically in young stellar associations. To date, we have discovered two previously unknown, large dimming events around the young stars RW Aurigae and V409 Tau. We attribute the dimming of RW Aurigae to an occultation by its tidally disrupted disk, with the disruption perhaps resulting from a recent flyby of its binary companion. Even with the dynamical environment of RW Aurigae, the distorted disk material remains very compact and presumably capable of forming planets. This system also shows that strong binary interactions with disks can also influence planet and core composition by stirring up and mixing materials during planet formation. We interpret the dimming of V409 Tau to be due to a feature, possibly a warp or perturbation, lying at least 10 AU from the host star in its nearly edge-on circumstellar disk. 

\end{abstract}

\firstsection 
\section{Introduction}
Young Stellar Objects (YSOs) are typically surrounded by protoplanetary circumstellar disks where the gas and dust eventually evolve over time through a combination of mechanisms including accretion onto the star, dispersion by stellar winds and radiation, and coalescence into planets. Other features of the stellar system, such as stellar companions, magnetic fields, or radiative jets, can influence the size, mass, and composition of these disks. The evolution of protoplanetary disks and the transition of dust and gas into planets is not well understood and directly affects our understanding of the processes that created the thousands of planetary systems thus far discovered. Characterization of structure in protoplanetary disks could help explain the differences between planetary systems. 

One means to better constrain the size, mass and composition of these disks arises by observing a star being eclipsed by its circumstellar disk. Thus far, only a few of these events have been detected and analyzed in the literature. One well-known example of this phenomenon is $\epsilon$ Aurigae, a bright system that periodically dims every 26.1 years by $\sim$1 magnitude for a duration of almost two years (\cite[Carroll et al. 1991]{Carroll:1991}). The eclipse has been attributed to an eclipsing binary where the companion star has a small circumstellar disk (\cite[Kloppenborg et al. 2010]{Kloppenborg:2010}). Another example of a large dimming event caused by a disk eclipse occurred in 2007 around the pre-main sequence star 2MASS J14074792-3945427. This star dimmed by $\sim$4 magnitude for $>$50 days, with the dimming believed to be caused by an occultation by a circumplanetary disk that is similar to Saturn's ring system but much larger (\cite[Mamajek et al. 2012]{Mamajek:2012}). The advent of wide-field time domain surveys provides an excellent tool to search for rare eclipse events, depending on the coverage, cadence, and baseline of the survey. Using time-series photometric data from the Kilodegree Extremely Little Telescope (KELT) exoplanet survey, which has coverage of a large portion of the sky, we are searching for disk eclipsing events, specifically in young stellar associations. Our survey has already yielded two discoveries of previously unknown large dimming events toward the T Tauri stars RW Aur and V409 Tau.
 \vspace*{-0.5 cm}
\section{KELT}
The Kilodegree Extremely Little Telescope (KELT) project is a survey of bright stars ($V$ = 8-10), with the goal to detect  transiting exoplanets. The survey uses two telescopes, KELT-South (Sutherland, South Africa) and KELT-North (Sonita, Arizona), each with a $26^{\circ}$ $\times$ $26^{\circ}$ field of view that observes with a $\sim$15 minute cadence in a broad $R$-band filter (\cite[Pepper et al. 2007, 2012]{Pepper:2007, Pepper:2012}). The KELT data are reduced using a heavily modified version of the ISIS software package, described in \S2 of \cite[Siverd et al. 2012]{Siverd:2012}\footnote{Much of the reduction software is publicly available:  http://verdis.phy.vanderbilt.edu}. RW Aur is located in KELT-North Field 04, which is centered on $\alpha$ =  5hr 54m 14.466s, $\delta$ = $+31^{\circ}$ 44' 37''. KELT-North observed this field from October 10, 2006 to September 23, 2012,  obtaining 8,001 images. V409 Tau and AA Tau are both located in KELT-North Field 03, which is centered on $\alpha$ = 3h 58m 12s, $\delta$ = $59^{\circ}$ 32' 24''. KELT-North observed this field for 7 seasons from UT 2006 October 26 to UT 2013 January 9, obtaining $\sim$9100 images. 
 \vspace*{-0.5 cm}
\section{Occultation of the T Tauri Star RW Aurigae A by its Tidally Disrupted Disk}
\begin{figure}[ht]
\begin{center}
\includegraphics[width=3in]{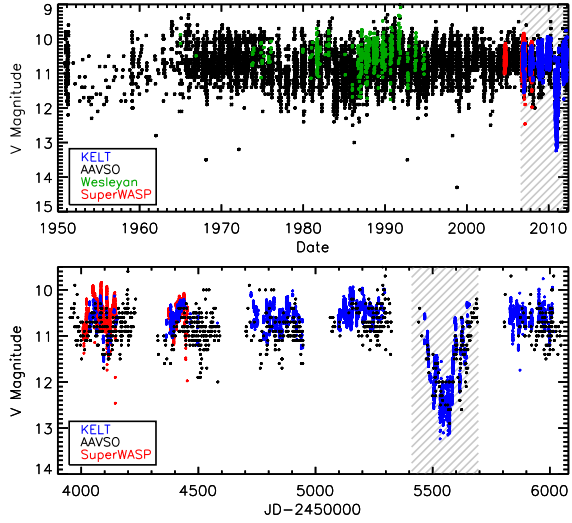} 
 \caption{(Top) AAVSO (Black), KELT-North (Blue), SuperWASP (Red) and the Wesleyan Van Vleck (Green) light curves of RW Aur from 1950 to 2012. The KELT and SuperWASP light curves do not resolve the A and B components. The shaded region in the upper plot corresponds to the six KELT seasons, which are shown in the bottom plot. (Figure reproduced from \cite[Rodriguez et al. 2013]{Rodriguez:2013}) }
   \label{fig:RWAUR}
\end{center}
\end{figure}

Results for RW Aur previously appeared in \cite[Rodriguez et al. 2013]{Rodriguez:2013} and are briefly summarized here. After being photometrically monitored for over 100 years (\cite[Beck \& Simon 2001]{Beck:2001}), with no large coherent dimming event observed during this period, the RW Aurigae system dimmed by $\sim$2 magnitudes for $\sim$180 days (Figure \ref{fig:RWAUR}). The dimming was observed by both the KELT-North telescope and the American Association of Variable Star Observers (AAVSO). Previous millimeter observations by \cite[Cabrit et al. 2006]{Cabrit:2006} suggest that the disk around the primary component, RW Aur A, had been tidally disrupted by a recent fly-by of RW Aur B. The material that was perturbed in this interaction appears to have coalesced into a large tidal arm wrapped around RW Aur A. We attributed the large dimming observed in 2010 to an occultation of RW Aur A by a portion of the tidally disrupted material. This hypothesis has been supported by hydrodynamical modeling of the RW Aurigae fly-by (\cite[Dai et al. 2015]{Dai:2015}).  This system illustrates how binary interactions can be crucial in sculpting circumstellar disks. Because of the occultation, we furthermore were able to show that the distorted tidal arm has retained a remarkably coherent structure.

\section{V409 Tau as Another AA Tau: Photometric Observations of Stellar Occultations by the Circumstellar Disk}
\begin{figure}[ht]
\begin{center}
\includegraphics[width=3.in]{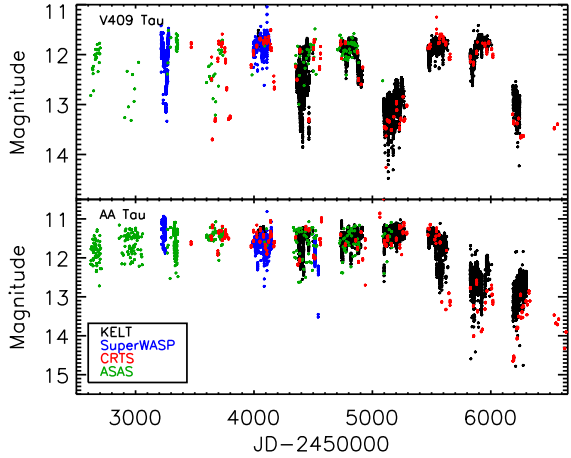} 
 \caption{KELT-North (Black), SuperWASP (Blue), CRTS (Red) and the ASAS (Green) light curves of V409 Tau (Top) and AA Tau (Bottom) from 2004 to 2013. (Figure reproduced from \cite[Rodriguez et al. 2015]{Rodriguez:2015})}
   \label{fig:V409}
\end{center}
\end{figure}

Results for V409 Tau previously appeared in \cite[Rodriguez et al. 2015]{Rodriguez:2015} and are briefly summarized here. Using the KELT-North data, we detected two large dimming events of the relatively unknown young star, V409 Tau. This system dimmed by 1.4 mag in early 2009 and then again in early 2012. We estimate the total duration of each event to be greater than 600 days. We also observed the well-studied dimming of AA Tau in 2011 (\cite[Bouvier et al. 2013]{Bouvier:2013}). AA Tau dimmed by $\sim$1.5 mag in 2011 and has not recovered. \cite[Bouvier et al. 2013]{Bouvier:2013} argued that the large dimming is caused by a feature in the close to edge-on circumstellar disk occulting the host star. Our spectral energy distribution (SED) analysis indicates that V409 Tau, like AA Tau, has a close to edge-on circumstellar disk. Using our observations of AA Tau as a direct comparison, we argue that these dimming events of V409 Tau are also the result of one or more features in the edge-on circumstellar disk, at a semi-major axis  of $>$10 AU, occulting the host star (\cite[Rodriguez et al. 2015]{Rodriguez:2015}).

 \vspace*{-0.7 cm}
\section{Conclusions}
Using the KELT survey, we are performing an all sky survey for YSOs presenting large dimming events, typically attributed to an occultation by a component within their circumstellar environment. Examples from the literature such as $\epsilon$ Aurigae and J1407, as well as our own discoveries, RW Aurigae and V409 Tau, illustrate the scientific value of large disk eclipses. The discovery of more such systems will permit a much wider range of studies to probe the size, structure, and composition of circumstellar disks. Our work is not only enhancing our knowledge of circumstellar environments and the early stages of stellar and planetary evolution, it provides a testbed and framework for the next generation of time domain photometric surveys. In particular, the Large Synoptic Survey Telescope (LSST) will increase the number of stars with long-baseline photometric observations by at least two orders of magnitude.  Furthermore, newly discovered systems with disk eclipses will be excellent targets for the NASA James Web Space Telescope observatory, which can probe the detailed environments of such systems.

\end{document}